\documentclass[twocolumn]{autart}

\usepackage{graphicx} 

\usepackage{amsmath,amsfonts,amssymb, enumitem}
\usepackage{soul, color}

\DeclareMathOperator{\trace}{trace}
\usepackage{epstopdf}

\begin{document}

\begin{frontmatter}

\title{Remote State Estimation with Privacy Against Active Eavesdroppers \thanksref{footnoteinfo}} 

\thanks[footnoteinfo]{A preliminary version of this work was presented at IFAC World Congress, 2023, Yokohama, Japan (see \cite{crimson2022remote}). Corresponding author Matthew Crimson.}

\author[First]{Matthew J.\ Crimson}
\author[First]{Justin M.\  Kennedy} 
\author[First]{Daniel E.\ Quevedo}

\address[First]{School of Electrical Engineering and Robotics, Queensland University of Technology, Brisbane QLD, 4000 Australia.
(e-mail: \{m.crimson, j12.kennedy,daniel.quevedo\}@qut.edu.au)}

\begin{keyword}              
Network Security, Privacy, Eavesdropping Attacks, Remote Estimation.         %
\end{keyword}

\begin{abstract} 
This paper considers a cyber-physical system under an active eavesdropping attack. A remote legitimate user estimates the state of a linear plant from the state information received from a sensor. Transmissions from the sensor occur via an insecure and unreliable network. An active eavesdropper may perform an attack during system operation. The eavesdropper intercepts transmissions from the sensor, whilst simultaneously sabotaging the data transfer from the sensor to the remote legitimate user to harm its estimation performance. To maintain state confidentiality, we propose an encoding scheme that is activated on the detection of an eavesdropper. Our scheme transmits noise based on a pseudo-random indicator, pre-arranged at the legitimate user and sensor. The transmission of noise harms the eavesdropper's performance, more than that of the legitimate user. Using the proposed encoding scheme, we impair the eavesdropper’s expected estimation performance, whilst minimising expected performance degradation at the legitimate user.
We explore the trade-off between state confidentiality and legitimate user performance degradation through selecting the probability that the sensor transmits noise.
Under certain design choices, the trace of the expected estimation error covariance of the eavesdropper is greater than that of the legitimate user.
Numerical examples are provided to illustrate the proposed encoding scheme. 
\end{abstract}
\end{frontmatter}
\section{Introduction}
In recent years, Cyber-Physical Systems (CPS) have grown rapidly with the ubiquitous rise of the Internet of Things. CPS integrate cyber-infrastructure and physical systems. These systems encompass large-scale, geographically dispersed systems, such as smart grids, intelligent transportation systems, critical infrastructure and  wearable medical devices \cite{ishii2022security}. Due to the tight integration of cyber and physical components in such systems, CPS are vulnerable to cyber-attacks. Cyber vulnerabilities can be exploited by malicious adversaries, which seek to disrupt CPS operations. Attacks on CPS may result in performance degradation or even system failure \cite{humayed2017cyber}. 

Security of a system is defined by availability, integrity, and confidentiality. Targeting these security goals are; denial-of-service (DoS), deception attacks, and eavesdropping. 
DoS attacks seek to compromise the availability of information via jamming communication channels. Deception attacks seek to target the integrity of a system by manipulating transmitted information. Eavesdropping attacks seek to intercept transmitted information, targeting confidentiality, \cite{cardenas2008secure}. In particular, the use of shared wireless mediums makes it difficult to shield transmissions from unintended recipients. Sophisticated attacks may combine elements to achieve goals, such as active eavesdropping. Active eavesdropping is where the eavesdropper can not only intercept transmissions but also invade the communication channel to degrade the wireless performance of the system \cite{ding2020remote}. 
This performance degradation may trigger the sensor to transmit data packets more frequently, or with higher power which increases the packet arrival rate at the eavesdropper. Therefore, estimation accuracy at the eavesdropper is increased. This style of attack differs from a DoS attack, as jamming is utilised to improve the eavesdropper performance as well, whereas a DoS attack seeks to degrade the legitimate user's performance, with no consideration to its own packet reception. In practical applications, active eavesdropping is achieved through pilot signal contamination \cite{6151778}.

There is a need to design estimation algorithms over wireless communication networks that account for the reliability of the wireless network, while also ensuring confidentiality of the transmitted information from an eavesdropper. Guaranteeing confidentiality is crucial for sensor and control data, which convey confidential information about the physical system state. This is detrimental to system operation as it violates sensor and user privacy, which may cause  economic losses or even threaten human survival. For example, in smart grids an eavesdropper can infer load taxonomy from the electricity consumption data provided by smart meters. The malicious adversary can infer daily schedules, and break into premises when no inhabitants are present \cite{10.1145/2382196.2382246}. In addition, with knowledge of how a system operates, an adversary may perform an attack with greater sophistication.

To ensure privacy against eavesdropping attacks, cryptography-based tools \cite{katz2020introduction} are at times used in practice, however, they introduce computation and communication overheads \cite{lee2010price} and therefore are only of limited use. Several works have developed encoding
schemes \cite{9762536,tsiamis2018state} and scheduling of transmissions \cite{8543618}, which are lightweight and avoid significant computation. These encoding techniques require knowledge of the legitimate estimator’s performance. Performance is shared through acknowledgment of successful packet receipts to the sensor.  However, utilising an acknowledgement channel is a point of vulnerability for eavesdropping and DoS attacks \cite{ding2020remote}, as experimentally demonstrated with the jamming of acknowledgments utilising commodity hardware \cite{klingler2019jamming}. Therefore, the motivation for this work is to design a low-complexity encoding scheme in a system without acknowledgments when an eavesdropper attack is occurring, by adapting transmissions to ensure the confidentiality of the state.

In this work, we consider a sensor transmitting the
state estimate of a stable, linear, dynamic system to a legitimate remote user over a random packet-dropping wireless network. 
A malicious adversary may perform an active eavesdropping attack during system operation and intercept the transmissions from the sensor, compromising state confidentiality. We propose an encoding scheme that is activated on the detection of an eavesdropper.
We present a Quickest Change Detection (QCD) scheme to detect the presence of an eavesdropper through monitoring channel outcomes.
QCD is a technique to detect an abrupt change in statistical properties of an observed process as quickly as possible after the change occurs, subject to a false alarm constraint \cite{BASSEVILLE1988309}.
Several formulations for QCD problems exist, which vary with respect to assumptions about the change time and optimality criteria. Bayesian formulations assume the change time is a random variable with a known geometric prior with independent and
identically distributed (i.i.d.) observations \cite{ford2020informativeness,unnikrishnan2011minimax}. With these assumptions, Shiryaev established the optimal stopping, which compared the change posterior probability to a threshold. 

In this work, we use QCD to detect the presence of an active eavesdropper, which activates the encoding scheme.
When an eavesdropper is detected, the sensor transmits noise or its local state estimate, based on a pseudo-random Bernoulli indicator that is pre-arranged at the legitimate user and sensor. Under this encoding scheme, the eavesdropper's estimation error covariance is made larger than that of the legitimate user. 
This work extends the results in \cite{crimson2022remote}. We extend \cite{crimson2022remote} from the scalar, first-order system to the vector case.
Additionally, no method was presented in \cite{crimson2022remote} for the detection of the eavesdropper. 

\section{Remote State Estimation with an Eavesdropper}
We consider the following dynamic system that is modelled as a linear, discrete-time invariant system
\begin{equation} \label{eq:lti}
 x_{k+1} = Ax_{k} + w_k,  \quad y_{k} = Cx_{k} + v_k
\end{equation}
where $x_k \in \mathbb{R}^n$ is the system state vector and $y_k \in \mathbb{R}^m$ is the measurement taken by the sensor at time $k \geq 0$. The dynamics is stable with $\rho(A) < 1$, where $\rho(.)$ is the spectral radius\footnote{The spectral radius is defined as the maximum absolute eigenvalue $\rho(.) = \max|\beta_i(.)|$ where $\beta_i$ is the $i$th eigenvalue.}. The process noise $w_k \in \mathbb{R}^n$ and the measurement noise $v_k \in \mathbb{R}^m$ are zero mean i.i.d. Gaussian random processes with $ \mathbb{E}[w_kw_k^{\intercal}] = Q$ with $\mathbb{E}[w_kw_l^{\intercal}] = 0$ and $ \mathbb{E}[v_kv_k^{\intercal}] = R$ with $\mathbb{E}[v_kv_l^{\intercal}] = 0$ and  $\mathbb{E}[w_kv_l^{\intercal}] = 0 \  \text{ for all } l \geq 0 \text{ \ where\ } k \neq l$. The initial state of the process $x_0$ is a Gaussian random variable with zero mean and covariance $\mathbb{E}[x_0x_0^{\intercal}] = \Sigma_0$. The covariances $Q, R, \Sigma_0$ are positive definite. The initial state is uncorrelated with $w_k$ and $v_k$ with $\mathbb{E}[x_0w_k^{\intercal}] = 0$ and $\mathbb{E}[x_0v_k^{\intercal}] = 0$. All system and noise parameters $(A, C, Q, R, \Sigma_0)$ are assumed to be \textit{public knowledge}, available to the sensor, legitimate user and eavesdropper. The pair $(A, C)$ is assumed to be observable and $(A, \sqrt{Q})$ is controllable.
\subsection{Local State Estimation and Open Loop Performance}
In current CPS frameworks, sensors are equipped with on-board computation and processes to improve the estimation performance of the system. Sensors process the collected measurements by executing recursive algorithms, enabled by advanced embedded systems-on-chip \cite{ding2017multi}. To emulate this sensor configuration, we consider the sensor configuration is computationally capable of optimal state estimation in real-time. The sensor locally estimates the system state $x_k$, based upon all measurements it has collected up to time $k$ using standard Kalman filtering recursions. The local estimate is transmitted wirelessly to a remote legitimate user. We denote $\hat{x}_{k}^{s}$ as the sensor's local Minimum Mean-Squared Error (MMSE) estimate of the system state, $x_{k}$, and $P_k^{s}$ as the corresponding estimation error covariance
\begin{equation*}
    \begin{split}
    \hat{x}_k^s &= \mathbb{E}[x_k|y_0, y_1,...,y_k] \\
        {P}_k^s &=\mathbb{E}[(x_k-\hat{x}_k^s)(x_k-\hat{x}_k^s)^{\intercal}|y_0, y_1,...,y_k].
    \end{split}
\end{equation*}
For any initial condition, the error covariance $P_k^s$ converges exponentially fast to some unique value $\Bar{P}$ \cite{anderson2012optimal}.
Without loss of generality, it is assumed the local state estimator has entered steady state operation and that $P_k^s = \Bar{P}$, where $\bar{P}$ is the unique solution to the discrete-time algebraic Riccati equation 
\begin{equation} \label{eq:pbar}
\begin{split}
    \bar{P} = &A\bar{P}A^{\intercal} + Q - (A\bar{P}A^{\intercal}+Q)C^{\intercal} \\&(CA\bar{P}A^{\intercal}C + CQC^{\intercal} + R)^{-1}C(A\bar{P}A^{\intercal} + Q) .
    \end{split}
\end{equation}
\subsection{Communication Model}
The legitimate user requires to reliably estimate the system state for the purposes of remote monitoring or control. To obtain this estimate, the sensor transmits an encoded packet $z_k \in \mathbb{R}^n$ of state information or noise. We define the formation of $z_k$ in Section \ref{sec:Transmissions at the Sensor}. The packet $z_k$ is transmitted over a wireless network to the legitimate user. However, the transmitted packets can also be received by an eavesdropper. We utilize a standard packet-based transmission utilized in network control problems. Denote the packet reception indicator outcome at the legitimate user by $\lambda_k \in \{0, 1\}$, where
\begin{equation*}
    \lambda_k = \begin{cases}
        1, \quad \textrm{if a packet is received at the legitimate user} \\
        0, \quad \textrm{if a packet dropout occurs.}
    \end{cases}
\end{equation*}
Denote the packet reception indicator outcome at the eavesdropper by $\lambda_k^e \in \{0, 1\}$, where
\begin{equation*}
    \lambda_k^e = \begin{cases}
        1, \quad \textrm{if a packet is received at the eavesdropper} \\
        0, \quad \textrm{if a packet dropout occurs.}
    \end{cases}
\end{equation*}
The channel outcomes for the legitimate user and eavesdropper are modelled as i.i.d. Bernoulli processes, independent of the initial state of the process and of the process noise. 
\subsection{Attack Model and Channel Qualities} \label{sec:attackmodel}
An eavesdropper seeks to compromise the confidentiality of the system state by eavesdropping on the transmitted state estimate from the sensor. At the same time, the eavesdropper invades the communication channel between the sensor and the legitimate user by attempting to block transmissions to degrade the estimation accuracy of the legitimate user. At some time $k = \Lambda$ where $\Lambda \geq 1 $  the eavesdropper performs an attack.

Let the channel quality be defined as the probability of packet dropout. In the absence of an eavesdropper attack,  the channel outcomes at the legitimate user $\lambda_k$ have a nominal packet dropout probability as $\gamma \in (0,1)$, where
\begin{equation*}
    \gamma  = \mathbb{P}[\lambda_k= 0] \quad k < \Lambda . 
\end{equation*}
We propose when the eavesdropper is performing an attack, the quality of the legitimate user's channel is degraded with a probability of packet dropout as $\bar{\gamma} \in (0,1)$, where
\begin{equation} \label{eq:legitchannel}
    \bar{\gamma} = \mathbb{P}[\lambda_k = 0] \quad k \geq \Lambda , 
\end{equation}
where $\gamma < \Bar{\gamma}$. When the eavesdropper is not performing an attack, the eavesdropper does not receive any information with $\lambda^e_k = 0$ when $k < \Lambda$, i.e., the eavesdropper does not have a communication channel.
When the eavesdropper is performing an attack, the probability of packet dropout for the eavesdropper's channel is defined as $\gamma^e \in (0,1)$, where
\begin{equation} \label{eq:eavesdropperchannel}
    \gamma^e = \mathbb{P}[\lambda_k^e = 0] \quad k \geq \Lambda.
\end{equation}
In our detection method we utilise the legitimate user's channel quality distributions to detect when an eavesdropper attack is occuring, detailed in Section \ref{sec:Eavesdropper Detection}.

\section{Problem of Interest}
As packet receipt acknowledgments may be vulnerable to adversaries, our objective is to design an encoding scheme that is activated on the detection of an eavesdropper, that does not rely on acknowledgements when the eavesdropper is detected. To characterise performance bounds under the encoding scheme for the eavesdropper's and legitimate user's estimation error covariance, we introduce ``open-loop performance''. We first define the solution to the discrete-time Lyapunov equation, which is expressed as an infinite sum in the case that $\rho(T) < 1$
\begin{equation} \label{eq:lyapsum}
    V = \sum_{j = 0}^{\infty}T^jU(T^{\intercal})^{j} ,
\end{equation}
where $V$ is the unique stabilising solution to
\begin{equation} \label{eq:lyap}
    V = TVT^{\intercal} + U .
\end{equation}
For convenience, we define the following notation to represent the solution to the Lyapunov equation
\begin{equation} \label{eq:lyapsol}
    V = L(T, U).
\end{equation}
If no information is received from the sensor at the legitimate user or eavesdropper, then open loop performance occurs. It is characterised by $ x^{OP}_k = \mathbb{E}[x_k] = 0$ and $P^{OP}_k = \mathbb{E}[x_kx_k^{\intercal}]$, where the asymptotic open loop estimate and corresponding estimation error covariance satisfy \cite{anderson2012optimal}
\begin{equation} \label{eq:openloop}
\begin{split}
 x^{OP} &\triangleq \lim_{k\to\infty}x_k^{OP}= 0 \\ P^{OP} &\triangleq \lim_{k\to\infty}P^{OP}_k=L(A,Q) .
 \end{split}
\end{equation} 
The encoding scheme is designed with no information of the legitimate user's current estimate, hence we aim to ensure that the trace of the legitimate user's expected estimation error covariance is upper bounded by the trace of open loop prediction. 
To ensure confidentiality, we desire that the trace of the eavesdropper's expected estimation error covariance is above  the trace of open loop prediction, i.e., the trace of the expected eavesdropper's estimation error covariance when using received packets is larger than if it had not used any packets in its estimation process. These conditions guarantee data confidentiality against an eavesdropper, whilst also ensuring the legitimate user has a reliable estimate.
\section{Encoding Scheme Design} \label{sec:Transmissions at the Sensor}
\begin{figure}
    \centering
    \includegraphics[width=8.4cm]{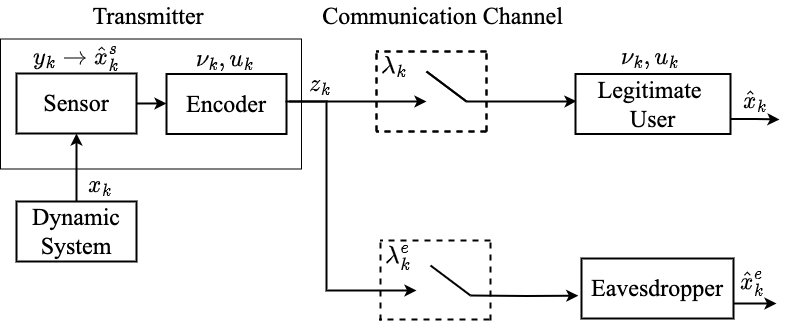}
    \caption{The sensor transmits $z_k$, an encoded packet formed by the indicator $u_k$ and alarm $\nu_k$. The alarm $\nu_k$ is formed at the legitimate user, indicating the presence of an eavesdropper and shared to the sensor. The indication to send noise $u_k$ is known to the legitimate user and sensor, and unknown to the eavesdropper. The packet is transmitted over a packet dropping network. The legitimate estimator and eavesdropper compute state estimates $\hat{x}_k$ and $\hat{x}_{k}^e$. }
    \label{fig:block}
\end{figure}
The considered remote estimation architecture is shown in Fig. \ref{fig:block} and consists of a sensor observing a dynamic system, a legitimate user, and an eavesdropper.
The sensor transmits packets of information over a wireless channel. 
To degrade the eavesdropper’s estimation performance, we desire to limit the amount of useful information it receives. 
To actively harm its estimation performance, we propose to send noise intermittently. 
\subsection{Encoding Scheme}
Our proposed strategy is to enact the encoding scheme, based upon the detection of an eavesdropper. We first introduce the decision variable $\nu_k \in \{0,1\}$, where
\begin{equation} \label{eq:nu}
\nu_k = 
\begin{cases} 
      0, &  \text{if eavesdropper has been detected}\\
      1, & \text{otherwise} . \\
\end{cases}
\end{equation}
The decision variable $\nu_k$ is formed at the legitimate user and shared with the sensor via an emergency acknowledgment channel. 
Once the sensor is alerted that an active eavesdropping attack is occurring, the sensor enacts the encoding scheme. We formulate eavesdropper detection and the design of $\nu_k$ using QCD in Section \ref{sec:Eavesdropper Detection}.

When an eavesdropper is detected, our proposed encoding scheme is to send either the sensor's state estimate $\hat{x}_k^s$ or noise $n_k$ by using a Bernoulli pseudo-random indicator $u_k \in \{0, 1\}$. The transmission of noise reduces transmissions of information to the legitimate user, reducing its performance. We pre-arrange the indicator $u_k$ between the legitimate estimator and sensor. This encoding scheme forms $z_k$ which is the transmitted packet from the sensor
\begin{equation} \label{eq:scheme}
z_k = 
\begin{cases} 
\hat{x}_k^s, & \text{ if } \nu_k = 1 \text{ or } (\nu_k, u_k) = (0, 1) \\
n_k, & \text{ if } (\nu_k, u_k) = (0,0). \\
\end{cases}
\end{equation}
The indicator $u_k$ informs what is transmitted at the sensor. In practical systems, random numbers are generated through pseudo-random algorithms. By agreeing upon an initial seed at the legitimate user and sensor, we pre-arrange the indicator sequence $u_k$. 

We choose the noise to have the same characteristics as the sensor's state estimate $\hat{x}_k^s$. The noise is designed such that a change in transmission cannot be easily detected at the eavesdropper, as it exhibits the same characteristics as the sensor's state estimate  $\hat{x}_k^s$  \cite{NAHA2023111147}. The noise $n_k$ is normally distributed, with zero mean $\mathbb{E}[n_k] = 0$ and covariance $\mathbb{E}[n_kn_k^{\intercal}] = \bar{P}$ uncorrelated with the state $x_k$ where $\mathbb{E}[x_kn_k] = 0$. 

We denote the probability of the sensor transmitting noise as $\mu \in (0,1)$, where
\begin{equation} \label{eq:muchoice}
    \mu \triangleq \mathbb{P}(u_k = 0).
\end{equation}
This probability constitutes the main design variable. It can be adjusted based on the dropout channel probabilities and the known process noise of the dynamics to balance the impact between degrading the legitimate user's performance and ensuring data confidentiality. We explore this trade-off in Section \ref{sec:design issues}.

Under this formulation, the goal is to decide how often to send noise such that data confidentiality is protected. The indicator is designed to be pseudo-random, such that the eavesdropper cannot easily discern patterns in transmission. Moreover, we require the sensor's state estimate to be received by the legitimate user intermittently to ensure its performance is not degraded by too many losses of state estimates from the sensor. Due to the pre-arrangement of the indicator $u_k$, the need for an acknowledgment channel is omitted when an eavesdropper is performing an attack. 
\subsection{Remote State Estimation}
We denote $\mathcal{I}_k$ as a collection of historical information at the legitimate user up to time $k$, with $\mathcal{I}_k \triangleq \{u_0, \lambda_0,\lambda_0z_0,...,u_k,\lambda_k,\lambda_kz_k\}$, noting that the legitimate user knows  the encoding scheme variable $u_k$. We define the legitimate user's own state estimate and the corresponding estimation error covariance as
\begin{equation} \label{eq:legituser}
    \begin{split}
    \hat{x}_k &\triangleq \mathbb{E}[x_k|\mathcal{I}_k] \\ P_k &\triangleq \mathbb{E}[(x_k-\hat{x}_k)(x_k-\hat{x}_k)^{\intercal}|\mathcal{I}_k],
    \end{split}
\end{equation}
which is based on all sensor data packets received up to time step $k$. We denote $\mathcal{I}^e_k$ as a collection of historical information at the eavesdropper up to time $k$ with $\mathcal{I}_k^e \triangleq \{\lambda_0^e, \lambda_0^ez_0,...,\lambda_k^e,\lambda_k^ez_k\}$. We define the eavesdropper's own state estimate and the corresponding estimation error covariance as 
\begin{equation} \label{eq:eaves}
    \begin{split}
    \hat{x}^e_k &\triangleq \mathbb{E}[x_k|\mathcal{I}_k^e]\\
    P_k^e &\triangleq \mathbb{E}[(x_k-\hat{x}_k^e)(x_k-\hat{x}_k^e)^{\intercal}|\mathcal{I}_k^e],
    \end{split}
\end{equation}
which is based on all sensor data packets received up to time step $k$. Note that the eavesdropper is unaware of the indication for the sensor to send noise $u_k$, Hence, the eavesdropper is deceived to incorporate noise $n_k$ into its state estimate.
\section{Remote State Estimation with an Eavesdropper} \label{sec:Remote State Estimation and Performance}
In an  unreliable, insecure network, the sensor's state estimate may not be received by the legitimate user, or the received packet may be noise due to the employed encoding scheme. To estimate the state of the system with intermittent observations, a second estimation scheme is employed at the legitimate user.

The legitimate user obtains the state estimate as follows: Once the sensor’s transmission arrives, and the sensor has not transmitted noise, the estimator synchronizes $\hat{x}_k$ with the optimal state estimate of the sensor $\hat{x}_k^s$. Otherwise, the legitimate user predicts $x_k$ based on the previous estimate using the system model in \eqref{eq:lti}. Thus, the state estimate at the legitimate user $\hat{x}_k$ \eqref{eq:legituser} satisfies
\begin{equation*}
    \hat{x}_k = \begin{cases}
        \hat{x}_k^s, \quad ~~~~~ \text{ if } (\lambda_k,u_k) = (1,1)\\
        A\hat{x}_{k-1}, ~~~ \text{ if } \lambda_k = 0 \text{ or }  (\lambda_k,u_k) = (1,0).
    \end{cases} 
\end{equation*}
The corresponding estimation error covariance $P_k$ satisfies
\begin{equation} \label{eq:estlegitimate}
    P_k = \begin{cases}
        \bar{P}, \quad ~~~~~~~~~~~~~~~ \text{ if }  (\lambda_k,u_k) = (1,1)\\
        AP_{k-1}A^{\intercal} + Q, ~ \text{ if } \lambda_k = 0 \text{ or } (\lambda_k,u_k) = (1,0).
    \end{cases} 
\end{equation}
It follows from the eavesdropper's estimation error covariance in \eqref{eq:estlegitimate} that when a packet arrives and it is the state estimate from the sensor, the legitimate user's estimation error covariance is that of the sensor's $\bar{P} $ (see \eqref{eq:pbar}). When a packet of noise arrives, or a packet dropout occurs, the legitimate user's estimation error covariance grows. 

The goal of the eavesdropper is to construct an estimate of the state of the system $x_k$  based on all sensor data packets received up to time step $k$. The eavesdropper is unaware of the detection variable $\nu_k$ \eqref{eq:nu}. The eavesdropper is additionally unaware of the randomised indicator to send noise $u_k$ \eqref{eq:scheme}. 
As the noise is designed to exhibit the same statistical characteristics as the sensor's state estimate, and the encoding scheme is unknown, the eavesdropper cannot distinguish $\hat{x}_k^s$ from $n_k$. 
This feature can be used to deceive the eavesdropper to incorporate noise $n_k$ into its state estimate. 
The eavesdropper obtains the state estimate as follows: Once the sensor’s transmission arrives, and the sensor has not transmitted noise, the eavesdropper synchronizes $\hat{x}^e_k$ with that of the sensor. If the sensor’s transmission arrives, and the sensor has transmitted noise, the eavesdropper synchronizes $\hat{x}^e_k$ with the transmitted noise $n_k$.  
If a packet dropout occurs, the eavesdropper predicts $x_k$ based on the previous estimate using the system model in \eqref{eq:lti}, such that the eavesdropper's state estimate $\hat{x}_k^e$ \eqref{eq:eaves} from the sensor's viewpoint satisfies
\begin{equation} \label{eq:eveasdropperestimation}
    \hat{x}_k^e = \begin{cases}
        \hat{x}_k^s, \quad ~~ \text{ if } (\lambda_k^e,u_k) = (1,1)\\
        n_k, \quad ~~ \text{ if } (\lambda_k^e,u_k) = (1,0)\\
        A\hat{x}_{k-1}^e, \text{ if } \lambda_k^e = 0. 
    \end{cases} 
\end{equation}
We characterise the eavesdropper's estimation error covariance in Lemma \ref{lemma:Estimation error covariance}.
\begin{lem} \label{lemma:Estimation error covariance}
The  estimation error covariance $P_k^e$ \eqref{eq:eaves} satisfies
\begin{equation} \label{eq:esteavesdropper}
    P_k^e = \begin{cases}
        \bar{P}, \quad ~~~~~~~~~~~~~~~\text{ if } (\lambda_k^e,u_k) = (1,1)\\
        P_n, \quad ~~~~~~~~~~~~~~ \text{ if } (\lambda_k^e,u_k) = (1,0) \\
        AP_{k-1}^eA^{\intercal} + Q, ~ \text{ if } \lambda_k^e = 0,
    \end{cases}
\end{equation}
where $P_n$ is 
\begin{equation} \label{eq:covuppereavesdropper}
    P_n = P^{OP} + \Bar{P},
\end{equation}
which is the estimation error covariance under receiving and utilising a packet of noise, as the packet is mistaken for the state estimate, with $P^{OP}$ defined in \eqref{eq:openloop}.
\begin{pf}
Firstly, consider the scenario that the eavesdropper successfully receives the state estimate from the sensor with $(\lambda_k^e,u_k) = (1,1)$, the eavesdropper's estimate is $\hat{x}_k^e=\hat{x}_k^s$ and the estimation error covariance is that of the sensor's (see \eqref{eq:pbar}) with $P_k^e = \bar{P}$. Secondly, in the scenario that a packet dropout occurs with  $\lambda_k^e = 0$, the eavesdropper predicts $\hat{x}_k^e$ with that of its previous estimate, with $\hat{x}_k^e=A\hat{x}_{k-1}^e$. The corresponding estimation error covariance is $P_{k}^e = AP_{k-1}^eA^{\intercal} + Q$. Thirdly, in the scenario that the eavesdropper successfully receives a packet of noise $(\lambda_k^e,u_k) = (1,0)$, the eavesdropper directly uses the packet as the state estimate $\hat{x}_k^e=n_k$. The estimation error covariance is
\begin{equation} \label{eq:covuppereavesdropperexpanded}
\begin{split}
P_n &= \mathbb{E}[(x_k-n_k)(x_k-n_k)^{\intercal}| \mathcal{I}_k^e] \\ &=  \mathbb{E}[x_kx_k^{\intercal}] + 2\mathbb{E}[x_kn_k] + \mathbb{E}[n_kn_k^{\intercal}] .
\end{split}
\end{equation}
These cases show the eavesdropper's estimation error covariance \eqref{eq:esteavesdropper}. By definition the transmitted noise is designed such that $\mathbb{E}[n_kn_k^{\intercal}] = \bar{P}$, uncorrelated with $x_k$ such that $\mathbb{E}[x_kn_k] = 0$.
As per \eqref{eq:openloop}, the open loop estimation error covariance is $\mathbb{E}[x_kx_k^{\intercal}] = P^{OP}$. Then from \eqref{eq:covuppereavesdropperexpanded}
\begin{equation*}
    P_n = \mathbb{E}[x_kx_k^{\intercal}] + 2\mathbb{E}[x_kn_k] + \mathbb{E}[n_kn_k^{\intercal}] = P^{OP} + 0 + \bar{P},
\end{equation*}
which shows the estimation error covariance under receiving a packet of noise \eqref{eq:covuppereavesdropper}.
\end{pf}
\end{lem}
Lemma \ref{lemma:Estimation error covariance} characterises the instantaneous worst-case performance for the eavesdropper under the encoding scheme.
It shows that if noise is transmitted often, we can drive the eavesdropper's estimation error covariance above open-loop prediction. 
However, greater transmissions of noise degrade the legitimate user's performance as well. Therefore, the probability of the sensor sending noise $\mu$ needs to be chosen to balance the impact between degrading the legitimate user's performance and protecting data confidentiality.

To quantify the estimation performance of the legitimate user and eavesdropper under the encoding scheme from a sensor viewpoint, we explore the mathematical expectation of the estimation error covariance. 
The use of the expectation allows us to compute the expected performance over the stochastic channel and encoding scheme. 
Particularly, due to the absence of acknowledgments when the eavesdropper is performing an attack, the channel outcomes are unknown at the sensor. Thus we use the expected estimation performance to design the sensor's encoding variable $\mu$. We additionally explore the steady-state behaviour of the expected estimation error covariance to quantify long-term performance under the encoding scheme as $k \rightarrow \infty$.

We present an analytic expression of the steady state expectation of estimation error covariance of the legitimate user in Lemma \ref{lemma:estimation error legitimate}  and of the eavesdropper in Lemma \ref{lemma:estimation error eavesdropper}. Our expressions depend on  the channel qualities \eqref{eq:legitchannel}, \eqref{eq:eavesdropperchannel}, process dynamics~\eqref{eq:lti}, and the encoding design variable \eqref{eq:muchoice}.

\begin{lem} 
\label{lemma:estimation error legitimate}
The steady-state expectation of the legitimate user's estimation error covariance satisfies
\begin{equation} \label{eq:esterrorlegitimatefull}
    \lim_{k\to\infty}\mathbb{E}[P_k] = (1-\bar{\gamma})(1-\mu){W} + (\bar{\gamma} + (1-\bar{\gamma})\mu){S} ,
\end{equation}
where $W$ and $S$ are solutions to the Lyapunov equation via \eqref{eq:lyap}, with
$W = L\left(\left(\sqrt{(\bar{\gamma} + (1-\bar{\gamma})\mu)}A\right), \Bar{P}\right)$
and
$S = L\left(\left(\sqrt{(\bar{\gamma} + (1-\bar{\gamma})\mu)}A\right), Q\right) $.
\end{lem}
\begin{pf}
The proof is included in Section \ref{sec:markov}. 
\end{pf}
\begin{lem} \label{lemma:estimation error eavesdropper}
The steady-state expectation of the eavesdropper's estimation error covariance satisfies
\begin{align} \label{eq:esterroreavesdropperfull}
    &\lim_{k\to\infty}\mathbb{E}[P_k^e]  \\
    &= (1-\gamma^e)(1-\mu)W^e + \gamma^eS^e + (1-\gamma^e)\mu H^e \nonumber
\end{align} 
where $W^e$ and $S^e$ and $H^e$ are solutions to the Lyapunov equation via \eqref{eq:lyap}, with 
$W^e = L(\sqrt{\gamma^e}A, \Bar{P})$, $S^e = L(\sqrt{\gamma^e}A, Q)$, and $H^e = L(\sqrt{\gamma^e}A, P_n)$.
\end{lem}
\begin{pf}
The proof is included in Section \ref{sec:markov}.
\end{pf}
\section{Markov Chain Model} \label{sec:markov}
The mathematical expectation of the legitimate user's and eavesdropper's estimation error covariance at time $k$ can be found by taking the sum of all possible channel and sensor transmission outcomes multiplied by the corresponding probability of that channel and sensor transmission realisation occurring. To find the mathematical expectation of the eavesdropper's and legitimate user's estimation error covariance, we introduce the following Markov chain models.
\subsection{Legitimate User's Markov Chain Model}
\begin{figure}
    \centering
    \includegraphics[width=8.4cm]{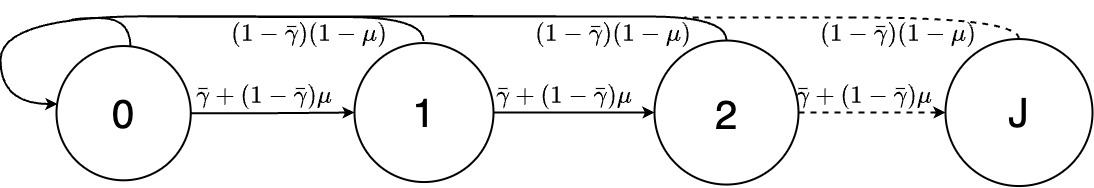}
    \caption{Markov chain model for the legitimate user. State $G_k = 0$ represents the legitimate user receiving the state estimate from the sensor. This event can be followed by infinite packet dropouts.}
    \label{fig:legitimateuserchain}
\end{figure}

Let $G_k$ be the state of a Markov chain taking values in the countably infinite set $G_k \in \{0,1,\dots\}$.
We define $G_k = 0$ as the state when the sensor's state estimate is received, and $G_k = j$ for $j > 0$ as the $j$th dropout.
By application of the legitimate user's channel quality \eqref{eq:legitchannel} and the probability of the sensor sending noise \eqref{eq:muchoice}, the transition probabilities $p_{ij} \triangleq \mathbb{P}[G_{k+1} = i |G_{k} = j]$ is characterised by
\begin{equation*}
    \mathbb{P}[G_{k+1} = i |G_{k} = j] = 
    \begin{cases} 
    (1-\bar{\gamma})(1-\mu), & \text{ if $i =  0$}, \\
    \bar{\gamma} + (1-\bar{\gamma})\mu, & \text{ if $i = j+1$}, \\
    0,  & \text{ otherwise}.
    \end{cases}
\end{equation*}
The Markov chain and its transitions are depicted in Fig. \ref{fig:legitimateuserchain}.
We define the transition matrix of the Markov chain as $\mathbf{P} = [p_{ij}]$.
Since all states communicate, the Markov chain is irreducible, aperiodic, and recurrent and has a stationary distribution $\pi$ \cite{bremaud2013markov}.
By solving the relation $\pi = \pi \mathbf{P}$ with $\sum_{j=0}^{\infty}\pi_j = 1$, the stationary distribution can be found as 
\begin{equation*} \label{eq:stationarylegitimate}
\begin{aligned}
\pi_j &= (\bar{\gamma} + (1-\bar{\gamma})\mu)^j(1-\bar{\gamma})(1-\mu).
\end{aligned}
\end{equation*}
Following the legitimate user's estimation error covariance in \eqref{eq:estlegitimate}, the conditional expectation of the estimation error covariance $P_k$ at the Markov chain state $G_k = j$ is 
\begin{equation*}
    \mathbb{E}[P_k|G_k = j] =  A^{j}\bar{P}(A^{\intercal})^{j} + \sum_{i=0}^{j - 1} A^iQ(A^{\intercal})^i .
\end{equation*}
Taking the limit as $k \rightarrow \infty$, then the expected estimation error covariance is
\begin{equation*}
\begin{split}
    &\lim_{k\to\infty}\mathbb{E}[P_k] =  \lim_{k\to\infty}\sum_{j=0}^{k} \mathbb{E}[P_k|G_k = j]\pi_j \\
    &=(1-\bar{\gamma})(1-\mu) \\&\quad \times \left(\sum_{j=0}^{\infty}(\bar{\gamma} + (1-\bar{\gamma})\mu)^j (A^{j}\bar{P}(A^{\intercal})^{j} + \sum_{i=0}^{j - 1} A^{i}Q(A^{\intercal})^{i} \right) \\
    &=(1-\bar{\gamma})(1-\mu) \\
    &\quad \times \sum_{j=0}^{\infty}(\sqrt{(\bar{\gamma} + (1-\bar{\gamma})\mu)}A)^{j}\bar{P}(A^{\intercal}\sqrt{(\bar{\gamma} + (1-\bar{\gamma})\mu)})^{j} \\
    &\quad+  (\bar{\gamma} + (1-\bar{\gamma})\mu)\\ &\quad \times \sum_{j=0}^{\infty} (\sqrt{(\bar{\gamma} + (1-\bar{\gamma})\mu} A)^{j}Q(A^{\intercal}\sqrt{(\bar{\gamma} + (1-\bar{\gamma})\mu})^{j} .\\ 
\end{split}
\end{equation*}
These infinite sums satisfy the solution to the
discrete-time Lyapunov equation (see \eqref{eq:lyapsum}) with \\  $\rho(\sqrt{(\bar{\gamma} + (1-\bar{\gamma})\mu} A) < 1$ and therefore can be written in the form of \eqref{eq:lyapsol}, where $W = L(\sqrt{(\bar{\gamma} + (1-\bar{\gamma})\mu)}A, \Bar{P})$ and $S = L(\sqrt{(\bar{\gamma} + (1-\bar{\gamma}))\mu} A,Q)$,
and can be written in the form of \eqref{eq:esterrorlegitimatefull}.
If $\bar{\gamma} = 1$ then no packets are received, or if $\mu = 1$ when only noise is sent then the legitimate user receives no packets containing the state estimate, as such the Markov chain does not return to state $G_k=0$ and no longer has a stationary distribution, and the expected steady-state estimation error covariance satisfies
\begin{equation*}
    \lim_{k\to\infty}\mathbb{E}[P_k] = P^{OP}, \text{ if $\bar{\gamma} = 1$ or $\mu = 1$}.
\end{equation*}
\subsection{Eavesdropper's Markov Chain Model}
\begin{figure}
    \centering
    \includegraphics[width=8.4cm]{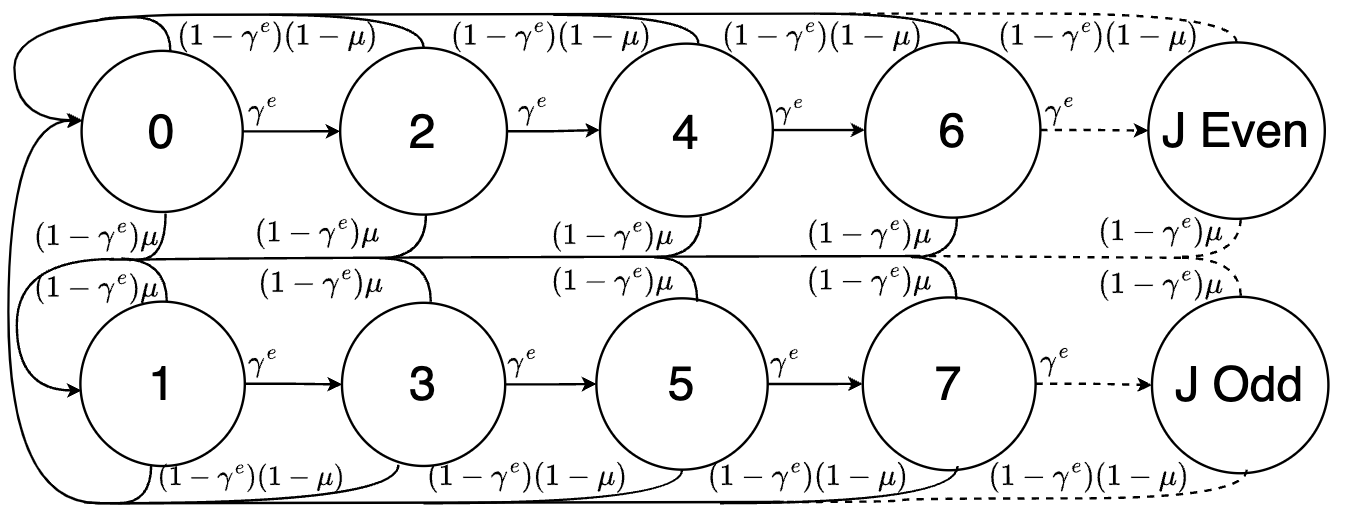}
    \caption{Markov chain model for the eavesdropper. State $G_k^e = 0$ represents the eavesdropper receiving the state estimate from the sensor. State $G_k^e = 1$ represents the eavesdropper receiving noise from the sensor. These two events can be followed by infinite packet dropouts.}
    \label{fig:eavesdropperchain}
\end{figure}

Let $G_k^e$ be the state of a countably infinite Markov chain taking values in the  set $G_k^e \in \{0,1,\dots\}$.
We define two key events, $G_{k}^e=0$ as the state when the sensor's state estimate is received, and $G_{k}^e=1$ as the state when noise is received by the eavesdropper.
These two events can be followed by an infinite number of packet dropouts.
Let $G_k^e=j$ for $j \geq 2$, be the dropouts after $G_\ell^e=0$ when $j$ is even, and $G_\ell^e=1$ when $j$ is odd, for some $\ell < k$.
The states $G_{k}^e=0$ and $G_{k}^e=1$ can be reached from any other state.
By application of  the eavesdropper's channel quality \eqref{eq:eavesdropperchannel} and the probability of the sensor sending noise \eqref{eq:muchoice}, 
the transition probability $p_{ij}^e \triangleq \mathbb{P}[G_{k+1}^e = i |G_{k}^e = j]$ is characterised by
\begin{equation*}
     \mathbb{P}[G_{k+1}^e = i |G_{k}^e = j] = 
    \begin{cases} 
    (1-\gamma^e)(1-\mu), & \text{ if $i = 0$,} \\
    (1-\gamma^e)\mu, & \text{ if $i =  1$,} \\
    \gamma^e,  & \text{ if $i = j + 2$}, \\
    0, & \text{ otherwise .}
    \end{cases} 
\end{equation*}
The Markov chain and its transitions are depicted in Fig. \ref{fig:eavesdropperchain}.
We define the transition matrix of the Markov chain as $\mathbf{P}^e = [p_{ij}^e]$.
Since all states communicate, the Markov chain is irreducible, aperiodic, and recurrent and has a stationary distribution $\pi^e$ \cite{bremaud2013markov}.
By solving the relation $\pi^e = \pi^e\mathbf{P}^e$ with $\sum_{j=0}^{\infty}\pi_j^e = 1$, the stationary distribution for all states are as follows 
\begin{align*}
    \pi_0^e &= (1-\gamma^e)(1-\mu), \\ \pi_1^e &= (1-\gamma^e)\mu, \\
    \pi_{j+2}^e &= (\gamma^e)\pi_j^e .
\end{align*}
Following the eavesdropper's estimation error covariance \eqref{eq:esteavesdropper}, the conditional expectation of the estimation error covariance $P_k^e$ at the Markov chain state $G_k^e = j$ is
\begin{align*}
    &\mathcal{P}_j^e = \mathbb{E}[P_k^e|G_k^e = j] \\
    & = \begin{cases}
        A^{\frac{1}{2}j}\Bar{P}(A^{\intercal})^{\frac{1}{2}j} + \sum_{i=0}^{\frac{1}{2}j - 1} A^{i}Q(A^{\intercal})^{i},&\text{ if $j$ even},\\ 
        A^{\frac{1}{2}(j-1)}P_n(A^{\intercal})^{\frac{1}{2}(j-1)} \\ + \sum_{i=0}^{\frac{1}{2}(j-1) - 1} A^{i}Q(A^{\intercal})^{i},&\text{ if $j$ odd} . 
    \end{cases}
\end{align*}
Taking the limit as time $k\rightarrow\infty$, then the expected estimation error covariance is
\begin{align*}
    &\lim_{k\to\infty}\mathbb{E}[P_{k}^e] = \lim_{k\to\infty}\sum_{j=0}^{k} \mathbb{E}[P_k^e|G_k^e = j]\pi_j^e \\
    &= \sum_{j=0}^{\infty} (\gamma^e)^j \pi_0^e \mathcal{P}_{2j}^e + \sum_{j=0}^{\infty}(\gamma^e)^j \pi_1^e \mathcal{P}_{2j + 1}^e 
    \end{align*}
    \begin{align*}
    &= (1-\gamma^e)(1-\mu) \sum_{j=0}^{\infty}(\gamma^e)^j\left(A^{j}\bar{P}(A^{\intercal})^{j} + \sum_{i=0}^{j - 1} A^{i}Q(A^{\intercal})^{i}\right)  \\  &\quad+ (1-\gamma^e)\mu\sum_{j=0}^{\infty}(\gamma^e)^j\left(A^{j}P_n(A^{\intercal})^{j} + \sum_{i=0}^{j - 1} A^{i}Q(A^{\intercal})^i\right) \\
    &= (1-\gamma^e)(1-\mu)\sum_{j=0}^{\infty}(\sqrt{\gamma^e}A)^{j}\bar{P}(A^{\intercal}\sqrt{\gamma^e})^{j} \\
    &\quad + (1-\mu)\gamma^e\sum_{j=0}^{\infty} (\sqrt{\gamma^e}A)^{j}Q(A^{\intercal}\sqrt{\gamma^e})^{j}  \\ 
    &\quad+ (1-\gamma^e)\mu\sum_{j=0}^{\infty}(\sqrt{\gamma^e}A)^{j}P_n(A^{\intercal}\sqrt{\gamma^e})^{j} \\ 
    &\quad+ \mu\gamma^e\sum_{j=0}^{\infty} (\sqrt{\gamma^e}A)^{j}Q(A^{\intercal}\sqrt{\gamma^e})^j .
\end{align*}
These infinite sums satisfy  the solution to the
discrete-time Lyapunov equation (see \eqref{eq:lyapsum}), with $\rho(\sqrt{\gamma^e}A) < 1$ and therefore can be written in the form of \eqref{eq:lyapsol}, where $W^e = L(\sqrt{\gamma^e}A,\Bar{P})$, $S^e = L(\sqrt{\gamma^e}A,Q)$, and $H^e = L(\sqrt{\gamma^e}A, P_n)$.
and can be written in the form of \eqref{eq:esterroreavesdropperfull}.
If $\gamma^e = 1$, then there are only dropouts and the Markov chain cannot return to state $G_k^e = 0$ or $G_k^e = 1$ for any $k > 0$ and no longer has a stationary distribution, and the expected steady-state estimation error covariance satisfies 
\begin{equation*}
    \lim_{k\to\infty}\mathbb{E}[P_{k}^e] = P^{OP} \text{ if $\gamma^e = 1$}.
\end{equation*}
\section{Encoding Design} \label{sec:design issues}

To provide a range of the encoding design variable $\mu$ \eqref{eq:muchoice}, we write the expectations of the estimation error covariances as a function of the encoding design variable. 
The expectation of the steady-state estimation error covariance of the legitimate user from Lemma \ref{lemma:estimation error legitimate} can be written as
\begin{equation} \label{eq:tracelegitimate} 
\begin{split}
    J(\mu) &= \trace\lim_{k\to\infty}\mathbb{E}[P_k] \\
    &= (1-\bar{\gamma})(1-\mu)\trace{W} \\ & \ + (\bar{\gamma} + (1-\bar{\gamma})\mu)\trace{S} ,
\end{split}
\end{equation} 
where $W$ and $S$ are functions of the encoding design variable $\mu$. Similarly, the expectation of the steady-state estimation error covariance of the eavesdropper from Lemma \ref{lemma:estimation error eavesdropper} can be written as
\begin{equation} \label{eq:treavesdropper}
\begin{split}
    J^e(\mu) &= \trace\lim_{k\to\infty}\mathbb{E}[P_k^e] \\
    &= (1-\gamma^e)(1-\mu)\trace W^e \\ & \ +\gamma^e\trace S^e + (1-\gamma^e)\mu \trace H^e
\end{split}
\end{equation}
Before providing a method to select the encoding design variable $\mu$, we observe the following monotonicity result.
\begin{lem} \label{lem:monotonicity}
Consider any eavesdropper channel quality $\gamma^e$ in $0<\gamma^e<1$ and
consider the encoding design variable \eqref{eq:muchoice} in $0 < \mu^\star < \mu < 1$. 
The trace of the expected steady-state estimation error covariance of the eavesdropper \eqref{eq:treavesdropper} is monotonically increasing in $\mu$ such that $J^e(\mu^\star) < J^e(\mu) $.
\end{lem}
\begin{pf}
    Consider $0 < \mu^\star < \mu < 1$. From the trace of the eavesdropper's expected steady-state estimation error covariance (see \eqref{eq:tracelegitimate}), the inequality $ J^e(\mu^\star) < J^e(\mu)$ is written as
    \begin{align*}
        &(1-\gamma^e)(1-\mu^{*}) \trace W^e + \\
        & \quad \gamma^e \trace S^e + (1-\gamma^e)\mu^{*}\trace H^e \\
        < &(1-\gamma^e)(1-\mu)\trace W^e + \\
        &\quad \gamma^e \trace S^e + (1-\gamma^e)\mu \trace H^e.
    \end{align*}
    By algebraic manipulation
    \begin{equation} \label{eq:eavesdropperincreasing}
    \begin{split}
        &\mu^{*}(1-\gamma^e)(\trace H^e - \trace W^e) \\ < &\mu(1-\gamma^e)(\trace H^e - \trace W^e). 
        \end{split}
    \end{equation}
    From \cite[Lemma 1]{192194}, let $P^{*} = W^e$, $\hat{P} = H^e$ with $Q^* = \Bar{P}$ and $\hat{Q} = P_n$. 
    As $P_n = P^{OP} + \Bar{P}$, it shows that $P_n > \Bar{P}$. Therefore $H^e > W^e$.  As $ \trace H^e> \trace W^e$, it shows that $\trace H^e - \trace W^e > 0$. 
    Therefore from the inequality \eqref{eq:eavesdropperincreasing}, it is shown that
    \begin{equation*}
        \mu^\star < \mu ,
    \end{equation*}
    which completes the proof. 
\end{pf}
Considering the choice of  $\mu$  in our encoding scheme, we desire to meet two conditions. 
Firstly, the trace of the legitimate user's expected steady-state estimation error covariance should be upper bounded by that of the trace of open-loop prediction, with $J(\mu) < \trace P^{OP}$.
This condition ensures that transmissions from the sensor remain informative to the legitimate user. 

Secondly, the trace of the eavesdropper’s expected steady-state estimation error covariance should be greater than the trace of open loop prediction, with $J^e(\mu) > \trace P^{OP}$.
This condition ensures that the confidentiality of the state is protected. 
The following Lemma provides an upper bound for the trace of the legitimate user's steady-state estimation error covariance. 
\begin{lem} \label{lem:JlessthanPOP}
Consider the encoding design variable $\mu < 1$  \eqref{eq:muchoice} and the legitimate user's channel quality $\Bar{\gamma} < 1$ \eqref{eq:legitchannel}. The trace of the expected steady-state estimation error covariance for the legitimate user \eqref{eq:tracelegitimate} is less than the trace of open-loop performance, with $J(\mu) < \trace P^{OP} $.
\end{lem}
\begin{pf}
    Consider the encoding design variable $\mu < 1$ and legitimate user's channel quality $\bar{\gamma} < 1$, and the eavesdropper has been detected with $\nu_k = 1$. Suppose that there is a non-zero probability that the sensor's state estimate is received at the legitimate user $\mathbb{P}[\lambda_k = 1, \nu_k=1, u_k = 0] > 0$. Then, from the Borel–Cantelli Lemma \cite{durrett2019probability}, a packet containing the state estimate from the sensor will arrive at the legitimate user with probability one for some time $k >0 $, that is, $\mathbb{P}[\lambda_k = 1, \nu_k=1, u_k = 0, \text{ for some } k>0] = 1$. The estimation error covariance of the legitimate user when the state estimate is received is $\Bar{P}$, noting that $\Bar{P} < P^{OP}$. Therefore, $ \lim_{k\to\infty}\mathbb{E}[P_k] <  P^{OP}$ if $\mu < 1$ and $\bar{\gamma} < 1$ which shows $J(\mu) < \trace P^{OP}$. This completes the proof.
\end{pf}
The following theorem provides a range for $\mu$ that protects data confidentiality, whilst keeping transmissions informative to the legitimate user.
\begin{thm} \label{thm:mu}
Consider the encoding design variable in \eqref{eq:muchoice} is chosen in the range $\mu^{OP}<\mu<1$, with $0 \leq(\Bar{\gamma}, \gamma^e) < 1$, given that
\begin{equation*}
    \mu^{OP} = \frac{\gamma^e(\trace S^e - \trace W^e) + \trace W^e - \trace P^{OP}}{(\gamma^e - 1)(\trace H^e - \trace W^e)} .,  
\end{equation*}
 then the following conditions hold
\begin{equation*}
    \begin{split}
        J(\mu) &< \trace P^{OP} \textrm{ and} \\
        \trace P^{OP} &< J^e(\mu) .
    \end{split}
\end{equation*}
\begin{pf} 
    To show the first case, consider the encoding design variable $\mu < 1$, then by Lemma \ref{lem:JlessthanPOP}, $J(\mu) < \trace P^{OP}$. 
    
    To show the second case, consider $\mu^{OP}$ as the probability of the sensor sending noise that ensures that $J^e(\mu^{OP}) = \trace P^{OP}$. Then from the trace of the eavesdropper's steady-state estimation error covariance in \eqref{eq:treavesdropper}, 
    \begin{equation} \label{eq:traceOPeavesdropper} 
    \begin{split} 
        \trace P^{OP} &= (1-\gamma^e)(1-\mu^{OP})\trace W^e \\ &+\gamma^e\trace S^e + (1-\gamma^e)\mu^{OP} \trace H^e . 
    \end{split}
    \end{equation}
    where $\mu^{OP}$ is found by re-arranging \eqref{eq:traceOPeavesdropper} in terms of $\mu^{OP}$. Consider the monotonicty result in Lemma \ref{lem:monotonicity}, and let $\mu^{OP} = \mu^\star$, then for any $\mu > \mu^{OP}$, $ J^e(\mu)>J^e(\mu^{OP})$, which shows that $ J^e(\mu)>\trace P^{OP}$. This completes the proof. 
\end{pf}
\end{thm}
In comparison with the state secrecy codes developed for stable systems \cite{tsiamis2018state}, the current method allows one to drive the eavesdropper's estimation error covariance above open-loop prediction through transmitting noise with $\mu > \mu^{OP}$. Thus, interceptions of packets are not informative to the eavesdropper's estimation process. 
Moreover, the eavesdropper intercepting packets is actively harming its own estimation performance. With the eavesdropper's expected estimation error covariance above open-loop prediction, the eavesdropper's performance is worse than if it did not utilise any packets at all from the sensor in its estimation process, i.e., if the eavesdropper only predicted the state estimate using the system model in \eqref{eq:lti}.
Since sending noise harms the legitimate user's performance as well, we choose $\mu < 1$ so the performance of the legitimate user remains upper-bounded by open loop prediction. This ensures that transmitted  packets from the sensor remain informative to the legitimate user for use in estimation.
\section{Relative Performance Simulation} \label{sec:relative}
\begin{figure}
    \centering
    \includegraphics[width=8.4cm]{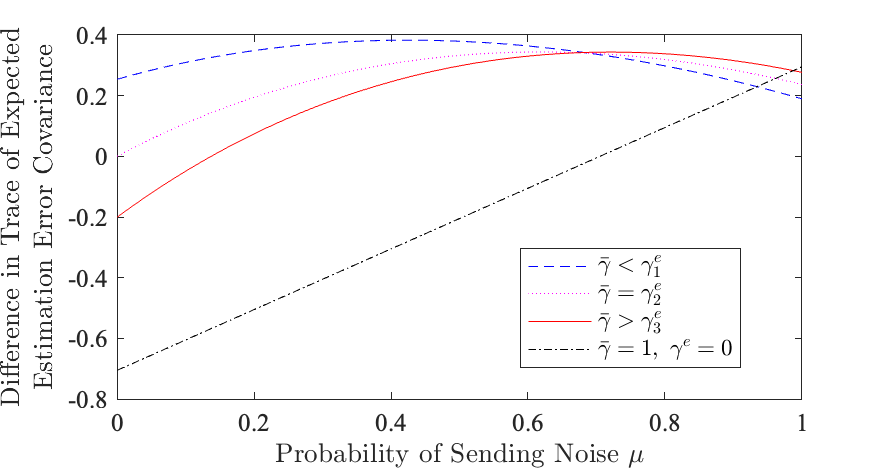}
    \caption{Comparison of the absolute difference in trace of the expected
estimation error covariance of the legitimate user and eavesdropper with four cases of channel qualities (worse, equal, better, much better).  Eavesdropper with worse channel quality in dashed blue, equal channel quality in dotted magenta, eavesdropper with better channel quality in solid red, and eavesdropper with perfect channel, legitimate user with no channel in dot-dashed black.}
    \label{fig:relperformance}
\end{figure}

In this section, we illustrate the performance of the encoding scheme using some examples. Consider the following second-order system \eqref{eq:lti} with parameters $A = \begin{bmatrix}
    0.5 & 0.1 \\ 0.4 & 0.6
\end{bmatrix}$, $C = I_2$,  $Q = R = 10^{-2}I_2$, where $I$ is the identity, and $\rho(A) = 0.7562 < 1$. 
This gives $\trace \bar{P} = 0.0114$, $\trace P_n = 0.0502$ and $\trace P^{OP} = 0.0388$.
Consider a legitimate user channel quality of $\bar{\gamma} = 0.5$ and four cases of eavesdropper channel qualities, with 
$\gamma^e_1 = 0.7$, $\gamma^e_2 = \bar{\gamma}$ $\gamma^e_3 = 0.3$. We also consider the case that $\gamma^e = 0, \Bar{\gamma} = 1$. 
Fig. \ref{fig:relperformance} shows the difference of the traces of the steady-state expected estimation error covariance between the eavesdropper and legitimate user $(J^e(\mu) -J(\mu))/J(\mu)$ against the encoding design variable $\mu$ for these four cases.
In the case that the eavesdropper's channel is worse than the legitimate user's, with $\bar{\gamma} < \gamma^e_1$  then the trace of the eavesdropper's estimation covariance is greater than the legitimate user's for $\mu \geq 0$ with $(J^e(\mu) -J(\mu))/J(\mu) > 0$. Using Theorem \ref{thm:mu}, $\mu$ is chosen in the range $0.7178< \mu < 1$. 
In the case that the eavesdropper's channel is equal to the legitimate user's with $\bar{\gamma} = \gamma^e_2$, then the trace of the eavesdropper's estimation covariance is greater than the legitimate user's for $\mu > 0$, with $(J^e(\mu) -J(\mu))/J(\mu) > 0$ if $\mu > 0$, and $(J^e(\mu) -J(\mu))/J(\mu) = 0$ if $\mu = 0$. Using Theorem \ref{thm:mu}, $\mu$ is chosen in the range $0.7125< \mu < 1$. 
In the case that the legitimate user's channel is worse than the eavesdropper's with $ \gamma^e_3 < \bar{\gamma}$, then the trace of the eavesdropper's estimation covariance is greater than the legitimate user's for $\mu > 0.14$. For $\mu \leq 0.14$, the trace of the legitimate user's expected estimation error covariance is greater than that of the eavesdropper's.  
Using Theorem \ref{thm:mu}, $\mu$ is chosen in the range $0.7074< \mu < 1$. 

We also present the case that $\Bar{\gamma} = 1$ and $\gamma^e = 0$, with the legitimate user not receiving any packets from the sensor, and the eavesdropper receives all packets transmitted from the sensor. If $\mu$ is chosen $\mu > 0.705$, then the trace of the eavesdropper's expected estimation error covariance is greater than the legitimate user's and above open loop prediction, even when the legitimate user is not receiving any transmissions from the sensor.

\section{Quickest Change Detection Formulation} \label{sec:Eavesdropper Detection}

To enact the proposed encoding scheme, the problem is to quickly detect the presence of an eavesdropper, utilising the channel outcomes $\lambda_k$ at the legitimate user. We present a Bayesian Quickest Change Detection (QCD) statistical test and a corresponding efficient computation method. 

Let $\lambda_{[0, k]} \triangleq \{\lambda_0,...,\lambda_k\}$ be shorthand for the legitimate user's channel outcome sequence. 
We model the instance of intrusion with a geometric distribution with $\epsilon_k = \mathbb{P}[\Lambda = k]= (1-\kappa)^{k-1}\kappa$ with $\kappa \in (0,1)$, where $\kappa$ is the probability that an eavesdropper attack can occur. 
This formulation is amenable in the framework of a standard QCD problem, where the goal is to detect a time in which the statistical properties of an observed process change \cite{NAHA2023111147}, which are the legitimate user's channel outcomes $\lambda_k$. 
Under the change description of the legitimate user's channel outcomes given in Section \ref{sec:attackmodel}, the joint probability mass function of the observed legitimate user packet outcomes is given by
    $p_\Lambda(\lambda_{[0,k]}) = \Pi_{i=1}^{\Lambda-1}(1-\gamma)\Pi_{j=\Lambda}^k(1-\bar{\gamma}) $,
given that $\Pi_{j=\Lambda}^k(1-\bar{\gamma}) = 1$ when $k < \Lambda$. 

Let $\mathcal{F}_k$ denote the filtration generated by $\lambda_{[0,k]}$ for $k \geq 0$.
We consider the probability space $(\Omega, \mathcal{F}, \mathbb{P}_{\Lambda})$, where $\Omega$ is the sample space of infinite sequences $\{\mathcal{F}_k: k \geq 0\}$ and $\mathcal{F} \triangleq \cup_{k=0}^\infty\mathcal{F}_k$ with the convention that $\mathcal{F}_0 = \{ 0, \Omega\}$ and $\mathbb{P}_\Lambda$ is the probability measure constructed from the joint
probability density $p_\Lambda(\lambda_{[0,k]})$ using Kolmogorov’s extension
theorem \cite{elliott2008hidden}.
Let us now construct a new probability measure $\mathbb{P}_{\epsilon}(G) = \sum_{k=1}^{\infty} \epsilon_k \mathbb{P}_k(G) $,
for all $G \in \mathcal{F}$. We will denote  the expectation operator
associated with $\mathbb{P}_{\epsilon}$ as  $\mathbb{E}_{\epsilon}$.
Our QCD problem is to quickly detect a change in the statistical properties of $\lambda_k$, which indicates that an active eavesdropper attack is occurring by designing a stopping time $\tau > 0$ with respect to the filtration $\mathcal{F}_k$. The stopping time minimises the following unconstrained (Bayes risk) optimisation problem of \cite{shiryaev1963optimum}, to trade-off average detection delay with a probability of false alarm as
\begin{equation} \label{eq:cost} 
    J(\tau) \triangleq c\mathbb{E}_{\epsilon}[(\tau - \Lambda)^+] + \mathbb{P}_{\epsilon}[\tau < \Lambda] , 
\end{equation}
where $(\tau - \Lambda)^{+} \triangleq \max(0, \tau - \Lambda)$ and $c$ is the penalty of each time step that alert is not declared after $\Lambda$.
\subsection{Eavesdropper Detection}
To enact the proposed encoding scheme, the legitimate user detects when an eavesdropper attack is occurring to form the decision variable $\nu_k$ \eqref{eq:nu}. Let $\mathbb{P}_{\epsilon}(k < \Lambda|\mathcal{F}_k )$ be the no-change posterior probability. Following \cite{unnikrishnan2011minimax}, the optimal stopping rule for the Shiryaev cost criterion \eqref{eq:cost} is
\begin{equation} \label{eq:stop}
    \tau^{*} = \inf\{k \geq 1: \mathbb{P}_{\epsilon}(k < \Lambda|\mathcal{F}_k ) \leq h\} ,
\end{equation}
given that $h \in (0,1)$ is a threshold chosen to control the probability of false alarms. At each time $k$, the no-change posterior $\mathbb{P}_{\epsilon}(k < \Lambda|\mathcal{F}_k )$ is calculated at the legitimate user using its channel outcome $\lambda_k$. An optimal stopping time is achieved under \eqref{eq:stop} to declare that an eavesdropper attack is occurring. This forms the decision variable $\nu_k$ \eqref{eq:nu}, which enacts the encoding scheme.
\begin{equation*}
\nu_k = 
\begin{cases} 
      0 &  \text{if $\tau^{*}$ has been declared}\\
      1 & \text{otherwise} . \\
\end{cases}
\end{equation*}
\subsection{Efficient Calculation of the No-Change Posterior}
We now present an efficient recursive method to compute the no-change posterior  $\mathbb{P}_{\epsilon}(k < \Lambda|\mathcal{F}_k )$. We define $\hat{M}_k^1 \triangleq \mathbb{P}_{\epsilon}(k < \Lambda|\mathcal{F}_k )$. Following \cite[Lemma 1]{ford2020informativeness}, given the sequence of legitimate user's channel outcome $\lambda_{k}$, the no-change posterior is provided by the scalar recursion
\begin{equation*}
    \hat{M}_k^1 = N_k(1-\kappa)(1-\gamma)\hat{M}_{k-1}^1 , 
\end{equation*}
with $\hat{M}_0^1 = 1$, and where $N_k$ is the normalisation factor 
\begin{equation*}
    N^{-1}_k =(1- \bar{\gamma} )+ (1-\kappa)(\bar{\gamma} - \gamma)\hat{M}^1_{k-1} . 
\end{equation*}
\section{Numerical Illustration with QCD} \label{sec:examples}
\begin{figure}
    \centering
    \includegraphics[width=8.4cm]{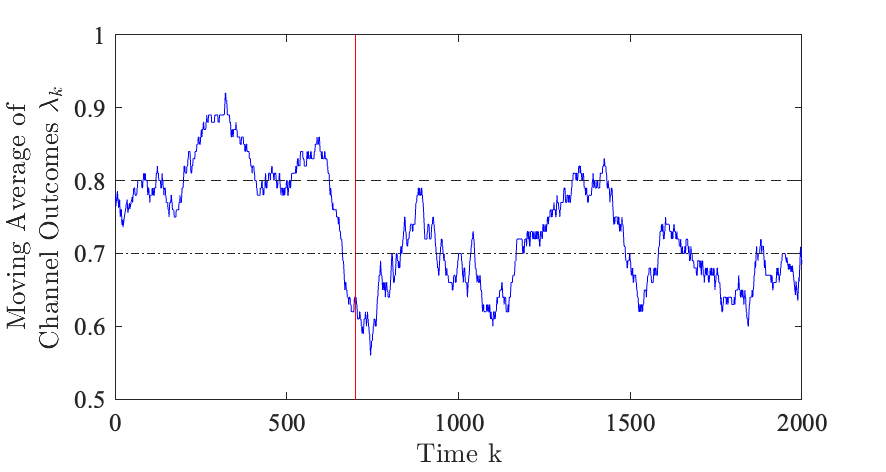}
    \caption{The moving average of the legitimate user's channel outcomes. The moving average is displayed with a blue solid line. The eavesdropper intrusion time is marked with a vertical red line. The theoretical mean of the channel outcomes at the legitimate user when an eavesdropper attack is not occurring is marked with a horizontal dashed black line, and when an eavesdropper attack is occurring is marked in a horizontal dot-dashed black line. }
    \label{fig:movav}
\end{figure}

\begin{figure}
    \centering
    \includegraphics[width=8.4cm]{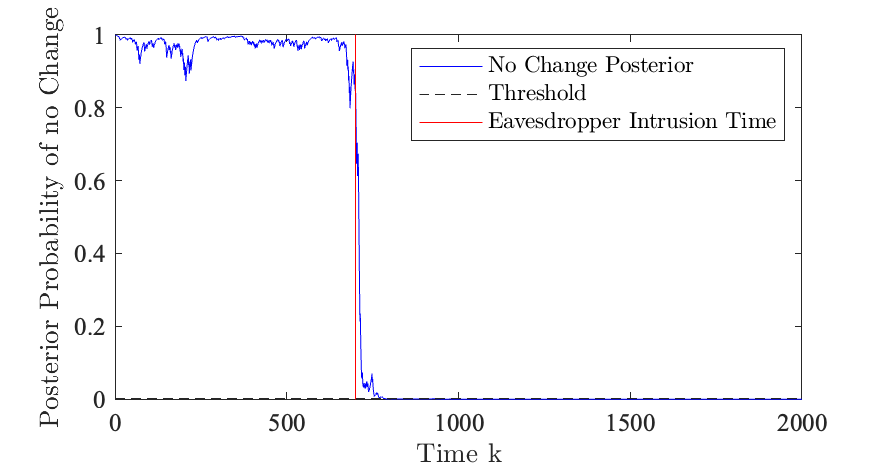}
    \caption{The no change posterior QCD test at the legitimate user against packet outcomes $\lambda_k$. The posterior probability of no change is marked with a blue line. The eavesdropper intrusion time is marked with a vertical red line. The zero false alarm threshold is marked by a horizontal black dashed line. }
    \label{fig:QCD}
\end{figure}

We show a numerical example to illustrate the results of QCD.  Consider the following second-order system detailed in Section \ref{sec:relative}. 
The eavesdropper intrusion time is $\Lambda = 700$. 
The nominal probability of channel dropout for the legitimate user is $\gamma = 0.2$. When the eavesdropper is performing an attack, the legitimate user's channel quality is degraded with $\bar{\gamma} = 0.3$. The eavesdropper's probability of channel dropout is $\gamma^e = 0.3$. 
To demonstrate the need for QCD in detecting the eavesdropper, we propose a moving average test over a window of length $N = 100$ computed on the channel outcomes at the legitimate user for a finite time horizon of $2000$ time steps. 
The moving average is displayed in Fig. \ref{fig:movav}. 
 The theoretical mean of the channel outcomes at the legitimate user under no attack is $1-\gamma = 0.8$, and when an eavesdropper attack occurring is $1-\Bar{\gamma} = 0.7$. While it is visually apparent that there is a decrease in the moving average, it is challenging to quickly detect the presence of the eavesdropper with certainty, where a naive user would declare a false alarm at time $k = 668$ in Fig. \ref{fig:movav}.
To quickly detect the presence of an eavesdropper, the proposed QCD test is applied. Fig. \ref{fig:QCD} displays the no change posterior calculated at the legitimate user for each time-step $k$. 
The threshold is chosen at the minimum value that achieves zero false alarms. This threshold for the given channel characteristics is $h = 3 \times 10^{-3}$ and was determined numerically via Monte-Carlo methods. The probability that an eavesdropper performs an eavesdropping attack is $\kappa = 5 \times 10^{-6} $. The legitimate user detects the presence of an eavesdropper at time $k = 779$, which is a detection delay of $79$ time steps. The posterior probability of no change reduces to a small value after the eavesdropper intrusion time. This indicates a strong probability of the presence of an active eavesdropper. Therefore at $k = 779$, a stopping time $\tau^{*}$ has been declared and the legitimate alarms the sensor, with $\nu_k = 0$. The indicator $\nu_k$ is shared with the sensor. The sensor then enacts the randomised scheme of transmitting noise or the state estimate.

\begin{figure}
    \centering
    \includegraphics[width=8.4cm]{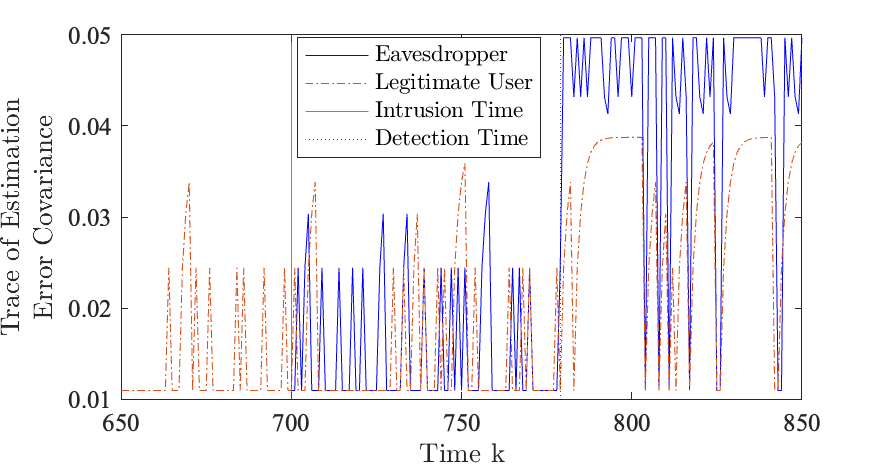}
    \caption{The trace of the estimation error covariance of the legitimate user, marked in solid blue, and eavesdropper, marked in dot-dashed orange for each time. The eavesdropper intrusion time is marked by a vertical solid red line. The eavesdropper detection time is marked by a vertical dotted black line. }
    \label{fig:esterrordirect}
\end{figure}

To show the performance over time with the detection scheme, we show the trace of the estimation error covariance of the legitimate user ($\trace P_k$, see \eqref{eq:estlegitimate}) and the eavesdropper ($ \trace P_k^e$, see \eqref{eq:esteavesdropper}) between the time window $k = 650$ and $k = 850$ in Fig. \ref{fig:esterrordirect}. The probability of sending noise $\mu$ is chosen to be $0.8$, which satisfies Theorem \ref{thm:mu}. 
Before the eavesdropper intrusion, the legitimate user is forming a state estimate and a corresponding estimation error covariance. The trace of the estimation error covariance increases when there are packet dropouts.
After the intrusion time, the eavesdropper performs an attack on the legitimate user's channel. There are more packet dropouts for the legitimate user and the trace of the legitimate user's estimation error covariance increases more often. The eavesdropper intercepts packets at this time and forms its own estimate of state and corresponding estimation error covariance.  
It is at time $k = 779$ that the eavesdropper is detected. The sensor enacts the encoding scheme at this time. Under successive packet dropouts or transmissions of noise from the sensor, the trace of the legitimate user's estimation error covariance is upper bounded by $\trace P^{OP}$. Under receiving a packet of noise, the trace of the eavesdropper's estimation error covariance is $\trace P_n$. 
\section{Conclusion} \label{sec:conclusion}
This paper has developed an encoding scheme that is activated on the detection of an active eavesdropper. 
We proposed that when the active eavesdropper is performing an attack, there is a statistical change in the dropout probability of the legitimate user's channel.
We propose a QCD statistical test to quickly detect the presence of an eavesdropper.
When an eavesdropper is detected, the sensor transmits either its state estimate or noise based upon a pseudo-random indicator.
The encoding scheme ensures the confidentiality of the state of a dynamic system, in a network environment without packet receipt acknowledgments when the eavesdropper is performing an attack.
Under our encoding design, the expected steady-state estimation error covariance of the eavesdropper is larger than that of the legitimate user. 
For certain design choices, the trace of the eavesdropper's expected steady-state estimation error covariance is larger than the trace of the open-loop estimation error covariance, whilst the trace of the legitimate user's expected steady-state estimation error covariance is smaller than the trace of open-loop estimation covariance. 
\bibliographystyle{plain} 
\bibliography{autosam}

\begin{thebibliography}{10}

\bibitem{anderson2012optimal}
{Anderson, B} and {Moore, J. B}.
\newblock {\em Optimal {F}iltering}.
\newblock Englewood Cliffs, N.J., USA: Prentice-Hall, 1979.

\bibitem{BASSEVILLE1988309}
{Basseville, M}.
\newblock Detecting {C}hanges in {S}ignals and {S}ystems—{A} {S}urvey.
\newblock {\em Automatica}, 24(3):309--326, 1988.

\bibitem{bremaud2013markov}
{Br{\'e}maud, P}.
\newblock {\em Markov {C}hains}.
\newblock New York, NY, USA: Springer-Verlag, 1999.

\bibitem{cardenas2008secure}
{Cardenas, A. A}, {Amin, S}, and {Sastry, S}.
\newblock Secure {C}ontrol: {T}owards {S}urvivable {C}yber-{P}hysical
  {S}ystems.
\newblock In {\em 28th International Conference on Distributed Computing
  Systems Workshops}, pages 495--500, Berkeley, California, 2008.

\bibitem{crimson2022remote}
{Crimson, M}, {Kennedy, J. M}, and {Quevedo, D. E}.
\newblock Remote {S}tate {E}stimation with {P}rivacy {A}gainst {E}avesdroppers.
\newblock {\em {IFAC} {W}orld {C}ongress, {Y}okohama, {J}apan
  arXiv:2211.13411}, 2023.

\bibitem{ding2017multi}
{Ding, K}, {Li, Y}, {Quevedo, D. E}, {Dey, S}, and {Shi, L}.
\newblock A multi-channel transmission schedule for remote state estimation
  under {DoS} attacks.
\newblock {\em Automatica}, 78:194--201, 2017.

\bibitem{ding2020remote}
{Ding, K}, {Ren, X.}, {Leong, A. S}, {Quevedo, D. E}, and {Shi, L}.
\newblock Remote {S}tate {E}stimation in the {P}resence of an {A}ctive
  {E}avesdropper.
\newblock {\em IEEE Transactions on Automatic Control}, 66(1):229--244, 2021.

\bibitem{durrett2019probability}
{Durrett, R}.
\newblock {\em Probability: {T}heory and {E}xamples}.
\newblock Cambridge university press, 2010.

\bibitem{elliott2008hidden}
{Elliott, R. J}, {Aggoun, L}, and {Moore, J. B}.
\newblock {\em Hidden {M}arkov {M}odels: {E}stimation and {C}ontrol},
  volume~29.
\newblock Springer Science \& Business Media, 2008.

\bibitem{ford2020informativeness}
{Ford, J. J}, {James, J}, and {Molloy, T. L}.
\newblock On the informativeness of measurements in {S}hiryaev’s {B}ayesian
  quickest change detection.
\newblock {\em Automatica}, 111:108645, 2020.

\bibitem{humayed2017cyber}
{Humayed, A}, {Lin, J}, {Li, F}, and {Luo, B}.
\newblock Cyber-{P}hysical {S}ystems {S}ecurity—{A} {S}urvey.
\newblock {\em IEEE Internet of Things Journal}, 4(6):1802--1831, 2017.

\bibitem{ishii2022security}
{Ishii, H} and {Zhu, Q}, editors.
\newblock {\em Security and {R}esilience of {C}ontrol {S}ystems}.
\newblock Springer, 2022.

\bibitem{katz2020introduction}
{Katz, J} and {Lindell, Y}.
\newblock {\em Introduction to {M}odern {C}ryptography}.
\newblock CRC press, 2020.

\bibitem{klingler2019jamming}
{Klingler, F} and {Dressler, F}.
\newblock Jamming {WLAN} {D}ata {F}rames and {A}cknowledgments using
  {C}ommodity {H}ardware.
\newblock In {\em IEEE Conference on Computer Communications Workshops}, pages
  1015--1016, Paderborn, Germany, 2019.

\bibitem{lee2010price}
{Lee, J}, {Kapitanova, K}, and {Son, S. H}.
\newblock The price of security in wireless sensor networks.
\newblock {\em Computer Networks}, 54(17):2967--2978, 2010.

\bibitem{8543618}
{Leong, A. S}, {Quevedo, D. E}, {Dolz, D}, and {Dey, S}.
\newblock Transmission {S}cheduling for {R}emote {S}tate {E}stimation {O}ver
  {P}acket {D}ropping {L}inks in the {P}resence of an {E}avesdropper.
\newblock {\em IEEE Transactions on Automatic Control}, 64(9):3732--3739, 2019.

\bibitem{9762536}
{Lücke, M}, {Lu, J}, and {Quevedo, D. E}.
\newblock Coding for {S}ecrecy in {R}emote {S}tate {E}stimation {W}ith an
  {A}dversary.
\newblock {\em IEEE Transactions on Automatic Control}, 67(9):4955--4962, 2022.

\bibitem{NAHA2023111147}
{Naha, A}, {Teixeira, A.M.H}, {Ahl{\'e}n, A}, and {Dey, S}.
\newblock Quickest detection of deception attacks on cyber–physical systems
  with a parsimonious watermarking policy.
\newblock {\em Automatica}, 155:111147, 2023.

\bibitem{192194}
{Poubelle, M.A}, {Bitmead, R.R}, and {Gevers, M.R.}
\newblock Fake {A}lgebraic {R}iccati {T}echniques and {S}tability.
\newblock {\em IEEE Transactions on Automatic Control}, 33(4):379--381, 1988.

\bibitem{10.1145/2382196.2382246}
{Rouf, I}, {Mustafa, H}, {Xu, M}, {Xu, W}, {Miller, R}, and {Gruteser, M}.
\newblock Neighborhood {W}atch: {S}ecurity and {P}rivacy {A}nalysis of
  {A}utomatic {M}eter {R}eading {S}ystems.
\newblock In {\em Proceedings of the 2012 ACM conference on Computer and
  communications security}, page 462–473, NY, USA, 2012.

\bibitem{shiryaev1963optimum}
{Shiryaev, A. N}.
\newblock On {O}ptimum {M}ethods in {Q}uickest {D}etection {P}roblems.
\newblock {\em Theory of Probability \& Its Applications}, 8(1):22--46, 1963.

\bibitem{tsiamis2018state}
{Tsiamis, A}, {Gatsis, K}, and {Pappas, G. J}.
\newblock State-{S}ecrecy {C}odes for {S}table {S}ystems.
\newblock In {\em 2018 American Control Conference (ACC)}, pages 171--177,
  Milwaukee, USA, 2018.

\bibitem{unnikrishnan2011minimax}
{Unnikrishnan, J}, {Veeravalli, V. V}, and {Meyn, S. P}.
\newblock Minimax {R}obust {Q}uickest {C}hange {D}etection.
\newblock {\em IEEE Transactions on Information Theory}, 57(3):1604--1614,
  2011.

\bibitem{6151778}
{Zhou, X}, {Maham, B}, and {Hjorungnes, A}.
\newblock Pilot {C}ontamination for {A}ctive {E}avesdropping.
\newblock {\em IEEE Transactions on Wireless Communications}, 11(3):903--907,
  2012.

\end{thebibliography}
\appendix
\end{document}